\documentclass[prb,twocolumn,showpacs]{revtex4}
\usepackage[dvips]{graphicx}

\begin{document}
\title{
Uniform Magnetic Order in a Ferromagnetic-Antiferromagnetic
Random Alternating Quantum Spin Chain
}

\author{
Tota Nakamura
}

\affiliation{
Department of Applied Physics, Tohoku University,
Aramaki aza Aoba 05, Aoba, Sendai, Miyagi  980-8579, Japan
}

\date{\today}

\begin{abstract}
An $S=1/2$ ferromagnetic (F) - antiferromagnetic (AF)
random alternating Heisenberg quantum spin chain model
is investigated in connection to its realization compound:
(CH$_3$)$_2$CHNH$_3$Cu(Cl$_x$Br$_{1-x}$)$_3$.
The exchange interaction bonds have alternating 
strong F-AF random bonds and weak uniform AF bonds.
Using the quantum Monte Carlo method we have found
that the excitation energy gap closes and the uniform AF order 
becomes critical in the intermediate concentration region.
This finding explains the experimental observation of the magnetic transition
by considering weak interchain interactions.
Present results suggest that the uniform AF order survives
even in the presence of randomly located ferromagnetic bonds.
This may be a new quantum effect.
\end {abstract}


\maketitle

\section{Introduction}

Quantum spin systems have been attracting wide interest both theoretically and 
experimentally.
The quantum effect sometimes produces a conclusion that is quite different from
our knowledge of the classical system.
Since the effect becomes remarkable in low dimensions,
various exotic phenomena have been found in 
one-dimensional quantum spin systems.
The Haldane state in an $S=1$ antiferromagnetic Heisenberg chain may be the
most famous example.\cite{haldane}
These theoretical systems can be realized experimentally.
New phenomena predicted theoretically can be observed in experiments,
and new findings from experiments can be explained theoretically.
The working together of theory and experiment in this field is producing
progress in condensed matter physics.

Randomness is another keyword in exotic phenomenon of quantum spin systems.
It sometimes brings order to disorder.
The appearance of a magnetic order due to impurity doping 
of a spin-disordered system is a typical example.\cite{hase,azuma}
There is a finite excitation energy gap above the nonmagnetic quantum 
ground state in the pure concentration compound.
Random impurity doping destroys the energy gap, and creates magnetic order.
A similar effect can be observed by destroying the energy gap 
by a uniform magnetic field.

Recently, 
Manaka et al.\cite{manaka3} found another example of randomness-induced
long-range order phenomena by mixing two spin-gapped compounds with nearly 
identical structures.
One compound is
(CH$_3$)$_2$CHNH$_3$CuCl$_3$ (abbreviated as IPACuCl$_3$), 
which realizes the $S=1/2$ ferromagnetic (F)-antiferromagnetic (AF) bond
alternation Heisenberg chain
with $J_{\rm strong}\sim$ 54 K (F) and $J_{\rm weak}\sim -23$ K (AF).
\cite{manaka-FAF1,manaka-FAF2}
Because of the strong F bonds, the ground state of this compound
is considered to be the Haldane state.
The other compound is
(CH$_3$)$_2$CHNH$_3$CuBr$_3$ (IPACuBr$_3$), 
which realizes the $S=1/2$ AF-AF bond alternation Heisenberg chain with
$J_{\rm strong}\sim -61$ K (AF) and $J_{\rm weak}\sim -33$ K (AF).
\cite{manaka-AFAF}
The ground state is the singlet dimer state.
These two compounds have different origins of the energy gap.

A magnetic phase transition is observed
in the intermediate concentration
$0.44 < x < 0.87$ of IPACu(Cl$_x$Br$_{1-x}$)$_3$
by measurements of the susceptibility and the specific heat.
Dependence of the susceptibility on the direction of the external 
field suggested that the order is antiferromagnetic.
The interesting point of this system is that the situation mentioned below is 
adverse to an existence of any uniform magnetic order.
On the other hand, the observed lines of experimental evidence suggest 
uniform antiferromagnetism.
Ferromagnetic bonds are randomly located in the mixed compound.
The staggered phase factor of the classical antiferromagnetic order
changes its sign randomly depending on the random location of the 
ferromagnetic bonds.
If the interchain interactions are uniformly antiferromagnetic 
(or ferromagnetic), there also exists randomly located frustration.

In this paper we show that uniform antiferromagnetic order
possibly appears in IPACu(Cl$_x$Br$_{1-x}$)$_3$.
Our results suggest a scenario of the appearance of the order.
The scenario can be expanded to explain a general impurity-induced long-range
order phenomenon.
A theoretical spin model is proposed in Sec. \ref{sec:model}.
We carry out quantum Monte Carlo simulations to this model.
The method is explained in Sec. \ref{sec:method}.
The results of the excitation energy gap, the string order parameter,
the magnetic susceptibility, and the staggered magnetization are
presented in Sec. \ref{sec:results}.
Section \ref{sec:conclusions} is devoted to conclusions.

\section{Model}
\label{sec:model}

We consider a theoretical model to explain the experiments on 
IPACu(Cl$_x$Br$_{1-x}$)$_3$.
From crystal structure analyses of pure compounds,
\cite{manaka-FAF1,manaka-FAF2,manaka-AFAF}
it is known that there are two Cl ions or two Br ions that bridge the 
magnetic Cu ions: Cu$<^{\rm Cl}_{\rm Cl}>$Cu or Cu$<^{\rm Br}_{\rm Br}>$Cu.
The copper ions are linked step-wise along the {\it c}-axis.
Exchange interactions within steps are strong, whereas those between
neighboring steps are weak.
The arguments are explained by the overlap of the 3d hole orbitals 
of Cu.\cite{manaka-FAF1,manaka-FAF2,manaka-AFAF}
The bridging angle differs between 
Cu$<^{\rm Cl}_{\rm Cl}>$Cu and Cu$<^{\rm Br}_{\rm Br}>$Cu
because of a different ion radius.
This difference may change the sign of exchange interactions
as well as their magnitude.
It is natural to consider that the strong bonds are more sensitive to
difference of angle.
Therefore, only strong bonds change the sign of interaction and
weak bonds remain antiferromagnetic in IPACuCl$_3$ and IPACuBr$_3$.

We neglect, for simplicity, a difference of magnitude of weak bonds
between IPACuCl$_3$ ($-23$~K) and IPACuBr$_3$ ($-33$~K) and set them 
a uniform antiferromagnetic interaction $-J$.
We also neglect a difference of magnitude of strong bonds of $54$~K and
$-61$~K in both compounds and set them $\pm 2J$.
In the mixed compound there is a configuration Cu$<^{\rm Cl}_{\rm Br}>$Cu.
It is not known what type of interaction it yields.
We assume that this is a strong interaction of $+2J$ or $-2J$
because the overlap of the hole orbitals is still large and
only the bridging angle differs slightly.
Then, the following ferromagnetic-antiferromagnetic random alternating
quantum spin chain model is considered.\cite{hida01a,totayukawa,hida01b,%
totaletter}
\begin{equation}
{\cal H}=-\sum_{i=0} ^{N-1}
J_{2i}
\mbox{\boldmath $S$}_{2i  }\cdot
\mbox{\boldmath $S$}_{2i+1},
+
J_{2i+1}
\mbox{\boldmath $S$}_{2i+1}\cdot
\mbox{\boldmath $S$}_{2i+2}
\end  {equation}
where $J_{2i}=\pm 2J$ indicates strong random bonds and
$J_{2i-1}=-J$ indicates weak uniform bonds.
The ferromagnetic bond concentration ratio on the strong bond is defined as $p$.
\begin{figure}
\begin{center}
\includegraphics[width=8.0cm]{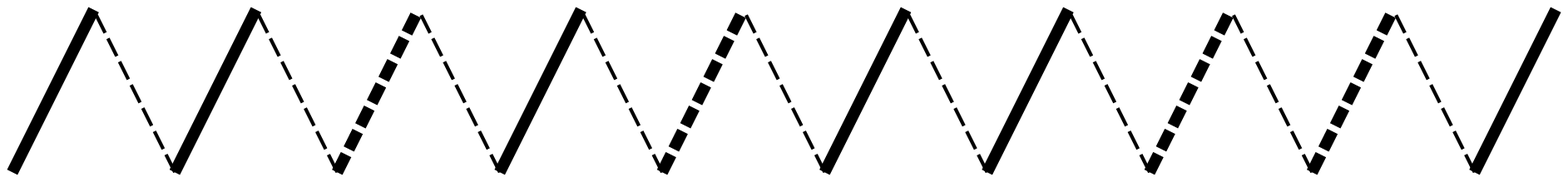}
\end{center}
\caption{
Theoretical model described in this paper.  Thick-solid lines and thick-broken
lines depict random ferromagnetic and antiferromagnetic interaction bonds, 
and thin broken lines depict uniform antiferromagnetic bonds.
}
\label{fig:model}
\end{figure}

Hida \cite{hida01a,hida01b} 
speculated that the mixed configuration Cu$<^{\rm Cl}_{\rm Br}>$Cu
yields a strong antiferromagnetic interaction ($-2J$) based on the discussion 
that the Haldane state is robust against randomness 
compared with the singlet dimer state.
This speculation is supported by recent experimental analyses,
\cite{manakajiba}
where dependence of the magnetic anisotropy constant and
the susceptibility maximum on atom concentration
are well explained by this speculation.
In this paper, we adopt this speculation.
Three bridging configurations are considered:
Cu$<^{\rm Cl}_{\rm Cl}>$Cu appearing with probability $x^2$, 
Cu$<^{\rm Br}_{\rm Br}>$Cu with $(1-x)^2$, and 
Cu$<^{\rm Cl}_{\rm Br}>$Cu with $2x(1-x)$.
Only the Cu$<^{\rm Cl}_{\rm Cl}>$Cu configuration is speculated to
yield ferromagnetic interactions.
Therefore, the Cl atom concentration $x$ is related to the ferromagnetic 
bond concentration $p$ by $p=x^2$.

We work only with this one-dimensional model.
If a magnetic order exhibits a critical behavior in one dimension, 
it is expected to be a long-ranged order
by finite three-dimensional interchain interactions.

Hyman and Yang \cite{hyman-y97} 
investigated a similar model wherein the alternate even bonds
are randomly AF and the other bonds are randomly F or AF.
If we apply the real-space renormalization group (RSRG) procedure \cite{ma79}
naively to the present model, it becomes an equivalent model.
When randomness is strong, a random singlet (RS) phase \cite{fisher94}
appears.
It is also suggested that the quantum Griffiths (QG) phase exists between the
Haldane phase and the RS phase.
However, it is not obvious how the classical magnetic order behaves in the
original model and where the phase boundary exists.
These points are made clear in this paper by using quantum Monte Carlo 
simulations.

\section{Method}
\label{sec:method}

Quantum Monte Carlo simulations are performed on the present model.
The model possesses randomness but there is no frustration.
The negative-sign problem does not appear.
We use a conventional world-line update algorithm.
A Trotter number of the Suzuki-Trotter decomposition is fixed as finite in the
simulation.
In equilibrium simulations data of finite Trotter numbers are extrapolated
to obtain a physical quantity in the infinite Trotter number limit where 
the original quantum system is recovered.
We calculate an excitation energy gap within this scheme.

Most simulations in this paper are based on the nonequilibrium relaxation (NER)
method.\cite{ner1,ner2,ner3,nonomura}
We start a simulation with a proper initial condition and observe how
physical quantities relax to the equilibrium values.
For example, an order parameter converges to a finite value in the 
ordered phase when the simulation starts from an ordered state.
It decays exponentially in the disordered phase, and
it decays algebraically in the critical phase.
The phase is identified by the behaviors of the relaxation of physical 
quantities.
It has been known that the method is particularly efficient in random 
and/or frustrated systems.\cite{shirahata,totasg1,totasgweakuniv,totaRbCoBr3}

The infinite size limit $L\to\infty$ is obtained beforehand in the NER scheme by
using a very large lattice and stopping the simulation before the finite-size
effect appears.
In the quantum Monte Carlo simulation it means that
the infinite Trotter number limit is also obtained beforehand.
A large $(d+1)$-dimensional system with a large Trotter number is prepared.
We stop the simulation before the finite-Trotter-number effect appears.
The obtained relaxation functions are equivalent to those in the infinite
Trotter number limit.
The quantum transition point can be estimated very accurately.\cite{totaqmc}
Nonomura \cite{nonomura}
argued that the transition point is dependent on the choice of the
Trotter number, and that the Trotter extrapolation is necessary
to obtain a transition point of the original quantum system.
This argument is not true.
We have developed a new quantum Monte Carlo algorithm of infinite Trotter
number \cite{totaqmc} and verified our argument.
It is also checked by comparing NER functions with different 
Trotter numbers.

\begin{figure}
\begin{center}
\includegraphics[width=8.0cm]{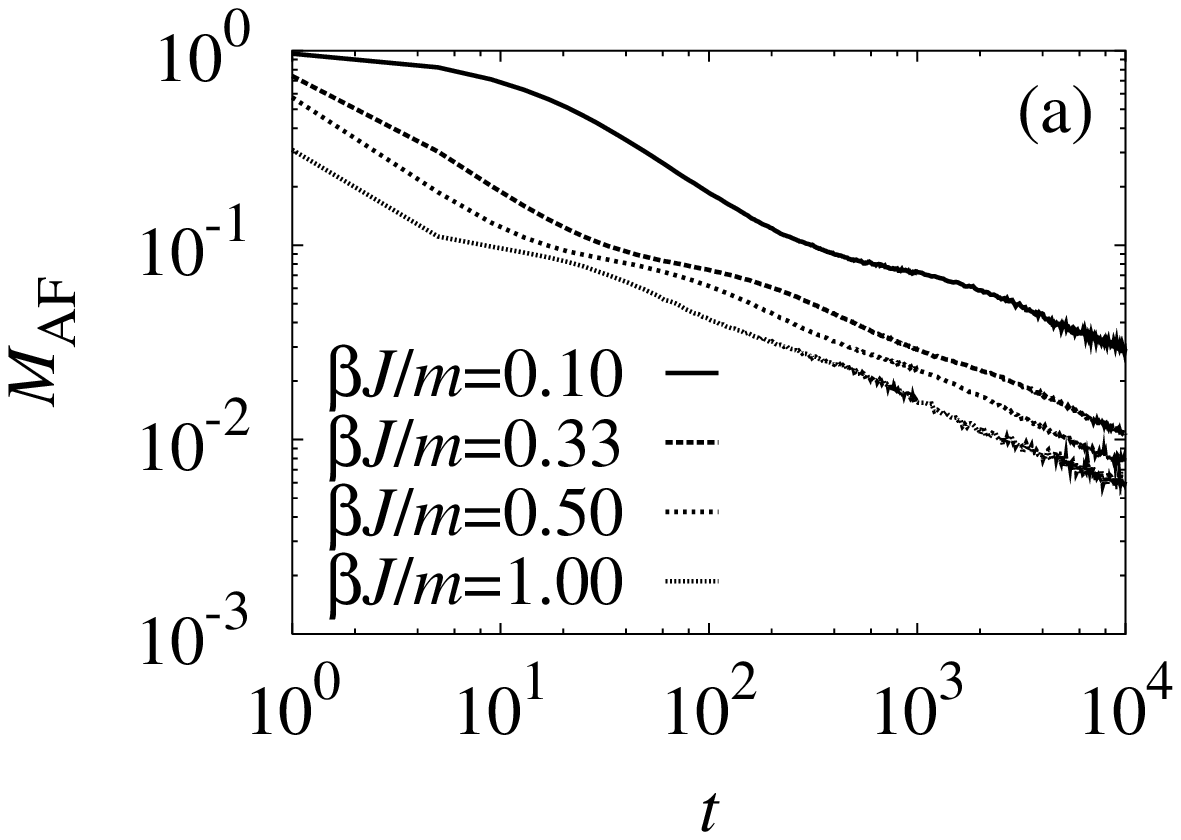}
\includegraphics[width=8.0cm]{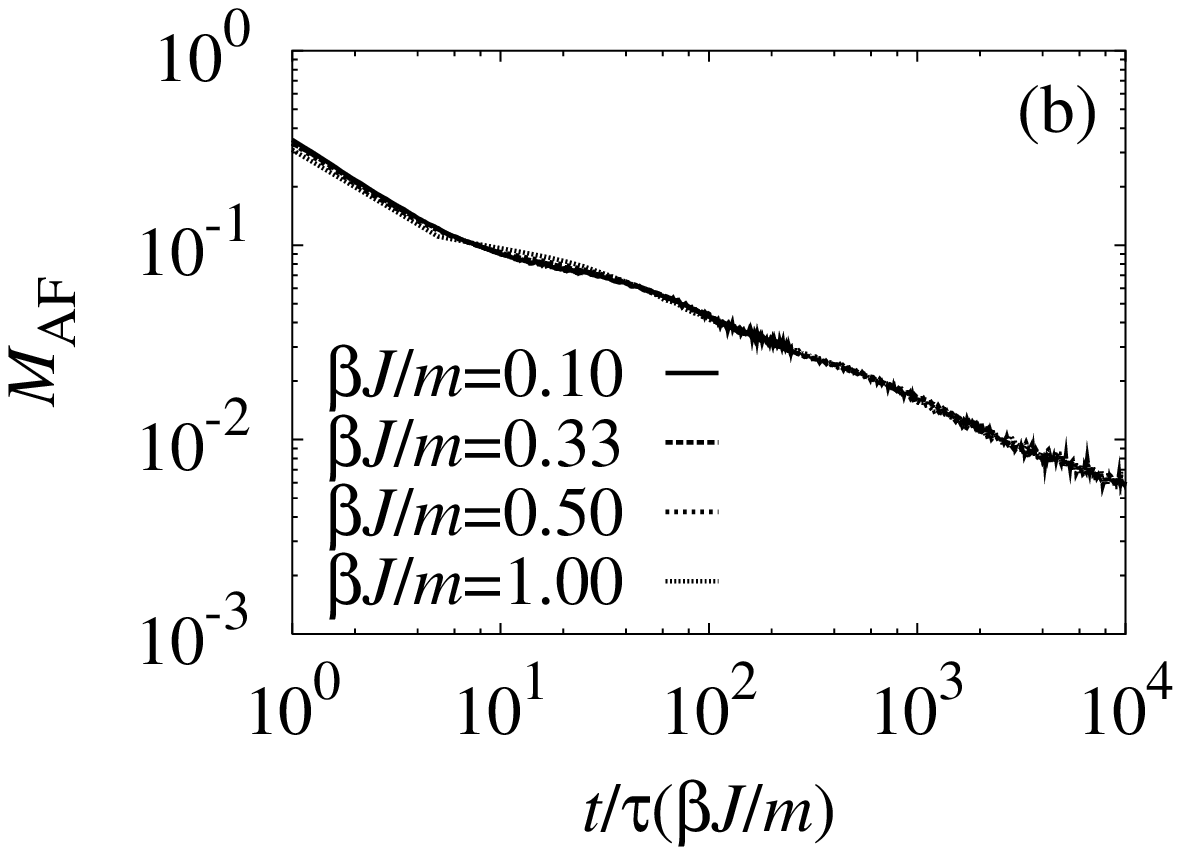}
\end{center}
\caption{
(a)
Relaxation functions of staggered magnetization ($M_\mathrm{AF}$) for
various choices of Trotter number $m$.
Inverse temperature $\beta J=50$ for $\beta J/m=0.10$ and $\beta J=500$ 
for the rest. 
The random bond concentration ratio $p=0.75$.
There is no temperature dependence within $10^4$ Monte Carlo
steps. Real-space system size is 644 spins for $\beta J/m=0.10$ and 
322 spins for the rest.
(b) If the Monte Carlo step is rescaled by $\tau(\beta J/m)$
of each Trotter number, the relaxation functions fall onto the same curve.
$\tau(0.10)=40$, 
$\tau(0.33)=4.2$, 
and
$\tau(0.50)=2.18$. 
}
\label{fig:trotter}
\end{figure}

Figure \ref{fig:trotter} shows relaxation functions of
staggered magnetization ($M_\mathrm{AF}$) when simulation starts from a
classical staggered state.
The raw relaxation curve is dependent on the Trotter number $m$ as shown
in Fig. \ref{fig:trotter}(a).
Relaxation becomes slow as the ratio $\beta J/m$ decreases.
This is called Wiesler freezing.
When the Monte Carlo step is rescaled by a correlation time of each 
Trotter number, the relaxation curves coincide with each other.
Therefore, the behavior of the relaxation function 
is independent of the Trotter number.
We can consider that the data of Fig. \ref{fig:trotter}(b) are the relaxation
function in the infinite Trotter number limit.
The value of $\beta J/m$ for NER simulations in this paper 
is fixed at $1/2$ unless otherwise stated.

\section{Results}
\label{sec:results}

\subsection{Excitation energy gap}

In this subsection we present results of the excitation energy gap.
It is investigated in order to make it clear where the gap closes 
by an introduction of the random bonds.
We adopt two different strategies.
One is direct observation of the energy gap.
It is calculated from the difference between an energy expectation value in a
subspace with total $S^z=0$ and that in a subspace with total $S^z=1$.
The other is a nonequilibrium relaxation of the local susceptibility
$\chi_\mathrm{loc}$ defined by:
\begin{equation}
T\chi_\mathrm{loc}
=\sum_i 
(\frac{1}{m}\sum_{j=1}^m S_{i,j}^z)^2
=\frac{2}{\Delta}
\sum_i 
\langle \psi_\mathrm{g}|S_i^z|\psi_\mathrm{ex}
\rangle^2.
\end  {equation}
Here, $j$ denotes a location in the Trotter direction in the 
$(d+1)$-dimensional lattice.
Subscript $i$ denotes a location in the real space direction.
Wave functions of the ground state and the first excited state are denoted by
$\psi_\mathrm{g}$ and $\psi_\mathrm{ex}$, respectively.
The excitation energy gap is denoted by $\Delta$.
Since the local susceptibility is proportional to the inverse of the excitation
gap, it diverges in the gapless phase and it converges to a finite
value in the gapful phase.
The change of behavior identifies the gapless-gapful transition point.
We have also performed a finite-time scaling analysis on the relaxation 
function of $T\chi_\mathrm{loc}$ and estimated the transition point.

\begin{figure}
\begin{center}
\includegraphics[width=8.0cm]{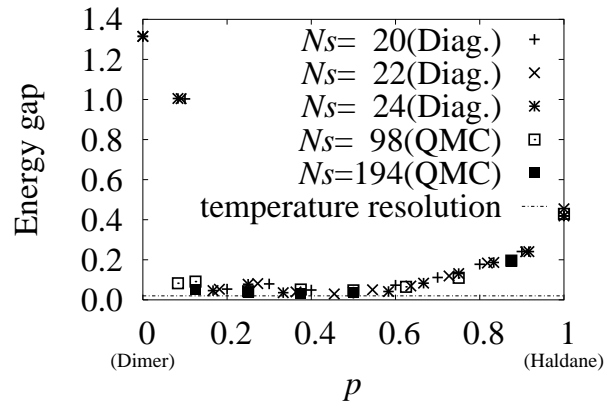}
\end{center}
\caption{Excitation energy gap estimated from an energy difference between 
the subspace with total $S^z=0$ and that with $S^z=1$.
}
\label{fig:gap}
\end{figure}
Figure \ref{fig:gap} shows the result of direct observation of the
energy gap.
When the spin number is small ($Ns \le 24$), we have done numerical 
diagonalization (Diag.) and estimated the numerically exact value for the 
energy gap.
Here, we have taken averages over all the possible random configurations
at each ferromagnetic bond concentration $p$.
When the spin number is large ($Ns \ge 98$), equilibrium quantum Monte Carlo
(QMC) simulations are carried out.
The temperature is $T/J=0.01$. Therefore, the energy gap resolution is 
an order of $\sim 0.02J$.
Infinite Trotter number extrapolations are performed using ten data between
$\beta J/m=0.38$ and $\beta J/m=1$.
Two thousand random bond configurations are taken.

We call a region $0.5<p<1$ a Haldane side hereafter in this paper.
At $p=1$ the Haldane state is realized in the ground state.
On the other side $0<p<0.5$ the region is called a dimer side.
This is because the dimer ground state is realized at $p=0$.
We call $p=0.5$ a fully-random point.

The energy gap on the Haldane side gradually decreases as random bonds are
introduced.
It seems that the gap closes at $p\sim 0.6$.
On the contrary, the gap on the dimer side suddenly decreases.
It suggests that the gap closes by a small randomness.
Diagonalization results at $p\sim 0.1$ are strongly affected by 
finite-size effects.

\begin{figure}
\begin{center}
\includegraphics[width=8.0cm]{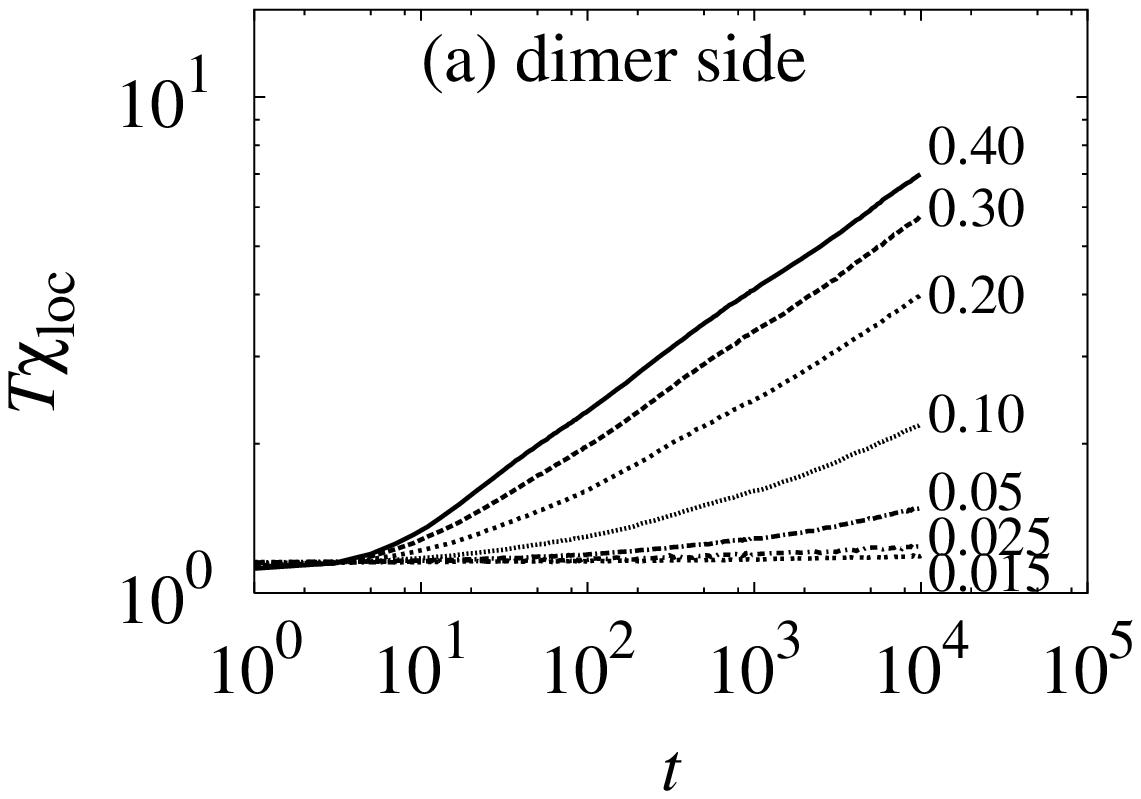}
\includegraphics[width=8.0cm]{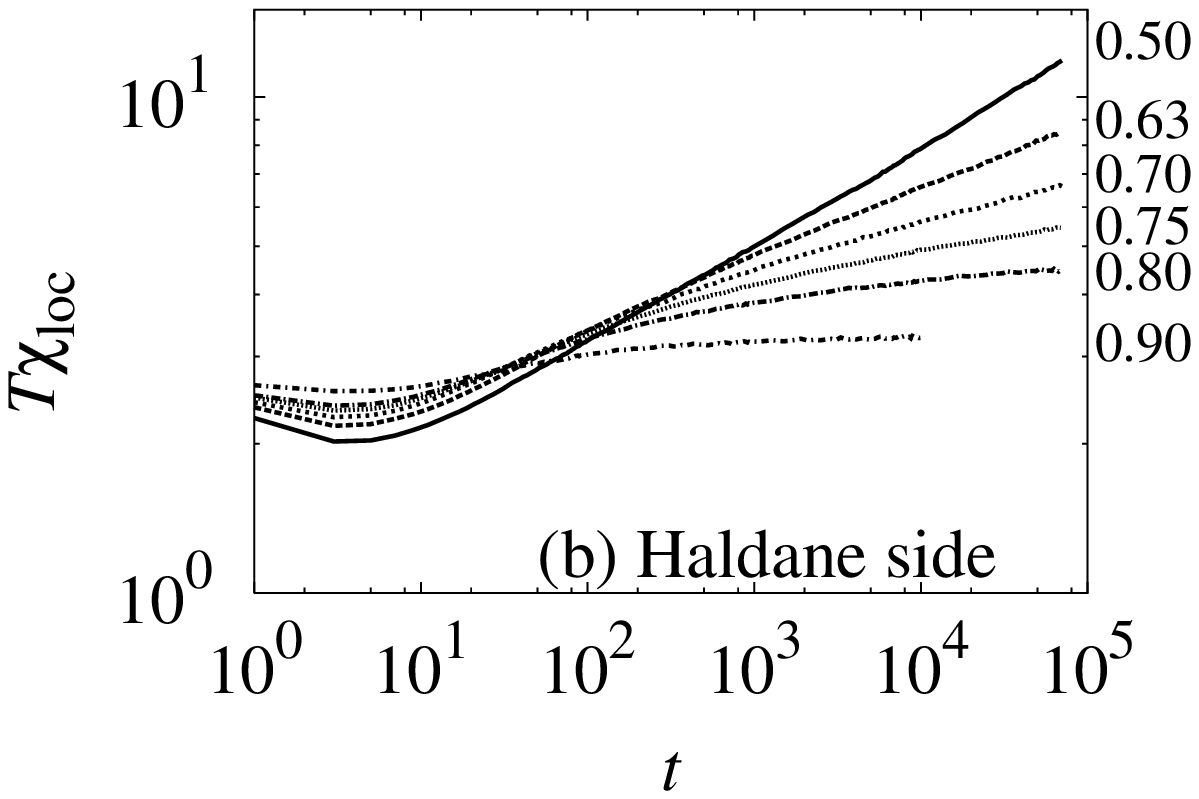}
\end{center}
\caption{Nonequilibrium relaxation of the local susceptibility.
A value of $p$ is as denoted beside each plot.
It exhibits diverging behavior in the gapless phase.
}
\label{fig:xloc}
\end{figure}
Figure \ref{fig:xloc} shows nonequilibrium relaxation plots 
for the local susceptibility.
Simulation starts from a gapful ground state in a pure concentration 
limit. On the dimer side the dimer ground state is prepared by running 
a simulation with the pure Hamiltonian at $p=0$.
By changing a random number sequence we have generated a different 
representation of the ground state for each random bond configuration.
On the Haldane side the Haldane ground state is prepared by a Hamiltonian
at $p=1$ in the same procedure.
The spin number is 322, the inverse temperature $\beta J=500$, and 
the Trotter number $m=1000$.  For $p=0.015$ the spin number is 802.
The number of random bond configurations is at least two thousand
and it is more than ten thousand near the transition point.

Gapless behavior is observed in the region of $0.025 < p < 0.75$.
There is a grey zone $0.63 < p \le 0.75$, where it is hard to identify
whether the relaxation function is diverging or converging.
Determination of the gapless-gapful transition point on the dimer side
is not settled in the present simulations.
There is a possibility that a relaxation of $p=0.015$ starts diverging 
behavior after $t=10^4$.
It is safe to note that the point is smaller than $p=0.025$.
It is also noted that the exponent of divergence (slope in the figure)
is almost independent of $p$ near $p=0.5$.
This suggests that the gapless phase near $p=0.5$ is characterized by some
universal exponent as in the random singlet phase.\cite{fisher94}

\begin{figure}
\begin{center}
\includegraphics[width=8.0cm]{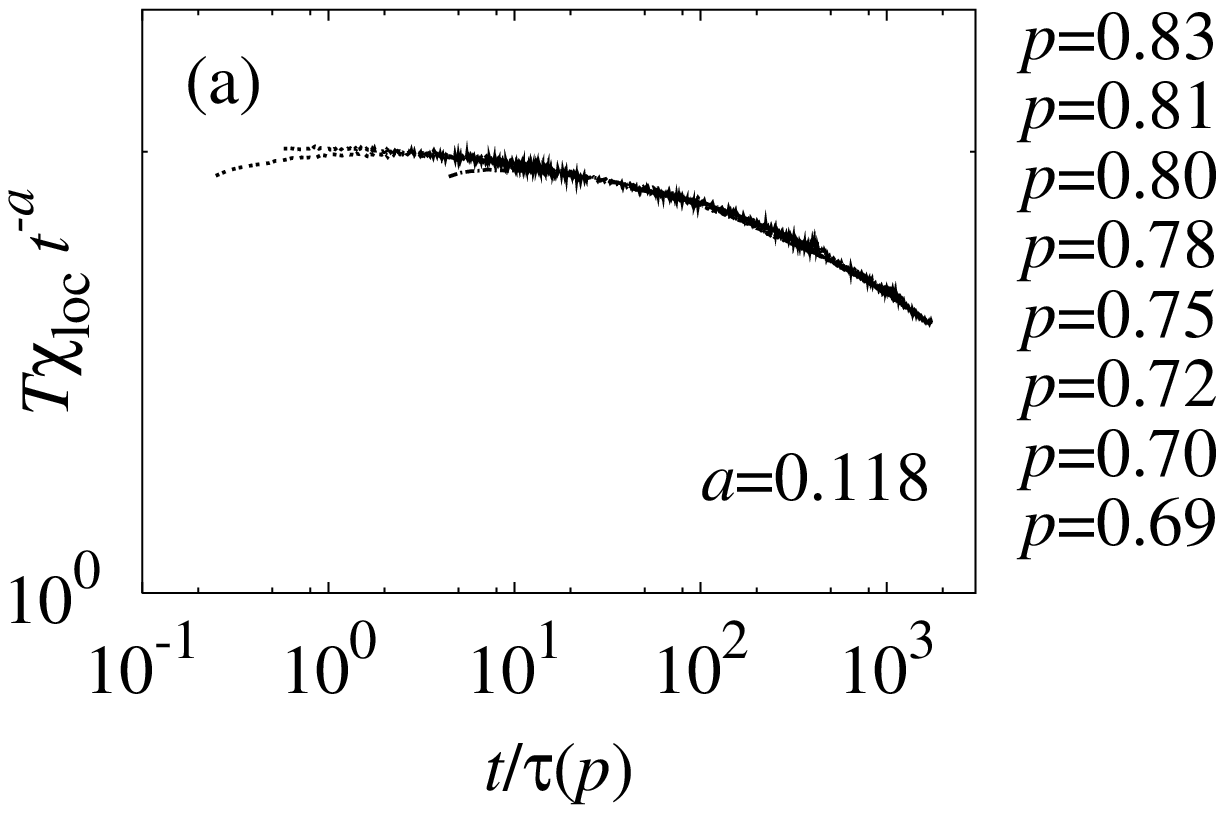}
\includegraphics[width=8.0cm]{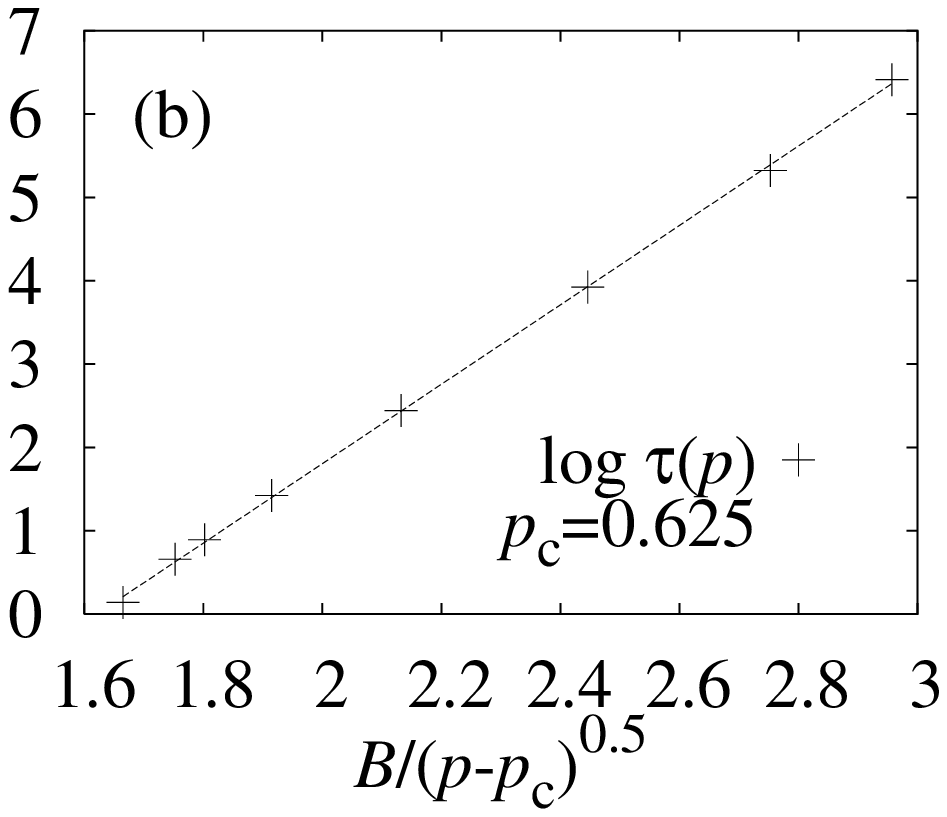}
\end{center}
\caption{Finite-time scaling of the local susceptibility on the 
Haldane side.
}
\label{fig:xlocscaling}
\end{figure}

In order to determine the gapless-gapful transition point on the Haldane side
we have performed a finite-time scaling analysis.\cite{ner3,shirahata}
We plot $T\chi_\mathrm{loc} t^{-a}$ versus $t/\tau(p)$ choosing 
$a$ and $\tau(p)$ properly so that all the relaxation data fall onto the 
same scaling function.
Figure \ref{fig:xlocscaling} (a) shows the result of scaling.
We have used data of eight different $p$ ranging $0.69 \le p \le 0.83$.
The correlation time $\tau(p)$ is supposed to diverge at the transition 
point following the KT singularity:
$\tau(p) \sim \exp[B/\sqrt{p-p_\mathrm{c}}]$.
It is shown in Fig. \ref{fig:xlocscaling} (b).
The best fit is achieved by a choice of transition point $p_\mathrm{c}=0.625$.
This procedure is a direct interpretation of the finite-size scaling analysis
of the KT transition in the $S=1/2$ alternating ferromagnetic chain
introduced by Yoshida and Okamoto.\cite{yoshidaokamoto}

Within the accuracy of the present simulations we can only note that
the gapful-gapless transition point on the dimer side 
is located somewhere below $p=0.025$.
Since this value is very small, we may consider that a small randomness
immediately destroys the excitation energy gap.
Therefore, the gapless phase is considered to exist in the region of
$ 0 < p < 0.625$.

\subsection{String order parameter}

The string order parameter is a good order parameter associated with the
gapless-gapful transition in a one-dimensional quantum spin system.
It reflects a hidden antiferromagnetic symmetry in the gapful phase.
On the Haldane side a string order parameter introduced by
den Nijs and Rommelse \cite{dennijs} is evaluated:
\begin{equation}
O_\mathrm{str} =
T_{i}^z\exp\left[i\pi \sum_{k=i+1}^{j-1}T_{k}^z
\right]
T_{j}^z,
\end  {equation}
where we have defined a bond spin 
$\mbox{\boldmath $T$}_{i}\equiv
 \mbox{\boldmath $S$}_{2i}+\mbox{\boldmath $S$}_{2i+1}$.
On the dimer side a string order parameter introduced by Hida \cite{hidastr}
is evaluated:
\begin{equation}
O_\mathrm{dim} =-4
S_{2i}^z\exp\left[i\pi \left(S_{2i+1}^z+\sum_{k=i+1}^{j-1}T_{k}^z
+S_{2j}^z\right) \right] S_{2j+1}^z.
\end  {equation}
These order parameters are considered to take finite values in the 
gapful phase.
It is argued that the string order parameter takes a finite value but the
excitation energy gap is zero in the quantum 
Griffiths phase.\cite{hyman-y97,yang-h2000}
Therefore, it is possible to identify this phase comparing the excitation 
gap results and the string order results.

\begin{figure}
\begin{center}
\includegraphics[width=8.0cm]{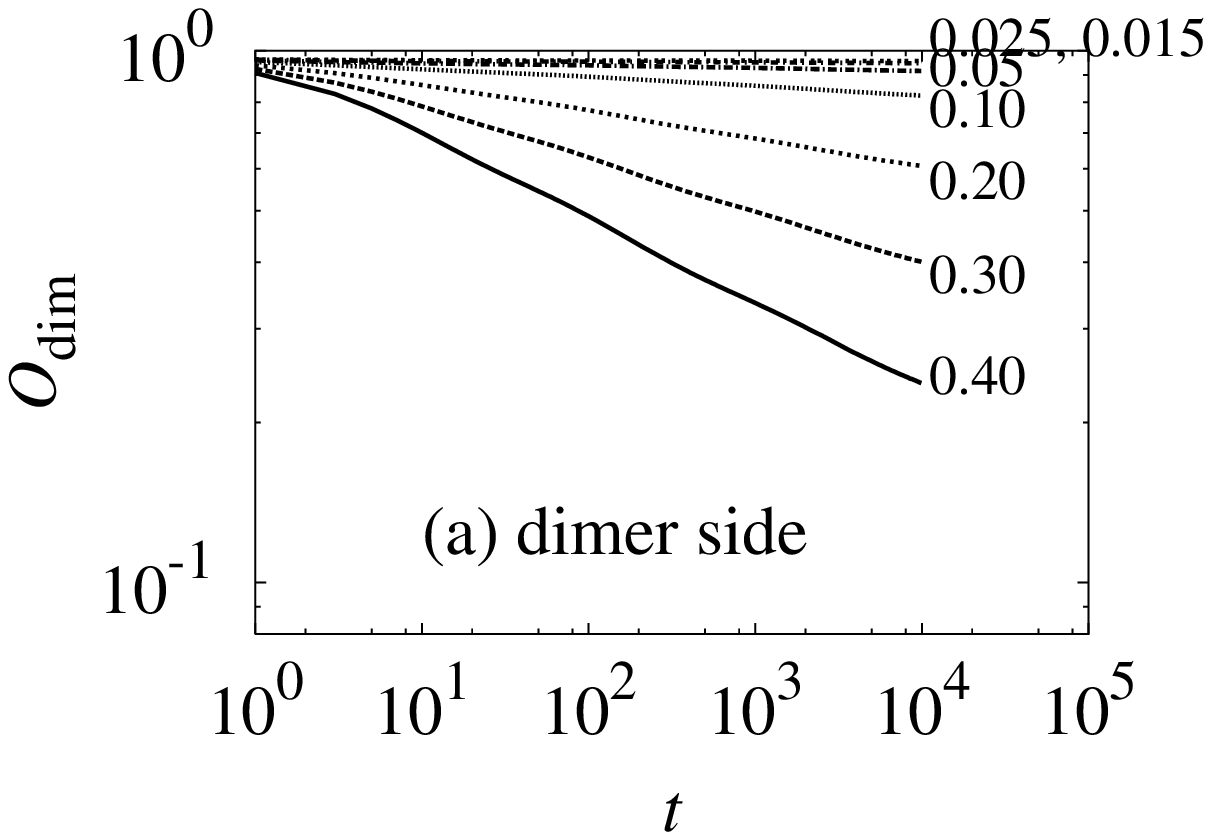}
\includegraphics[width=8.0cm]{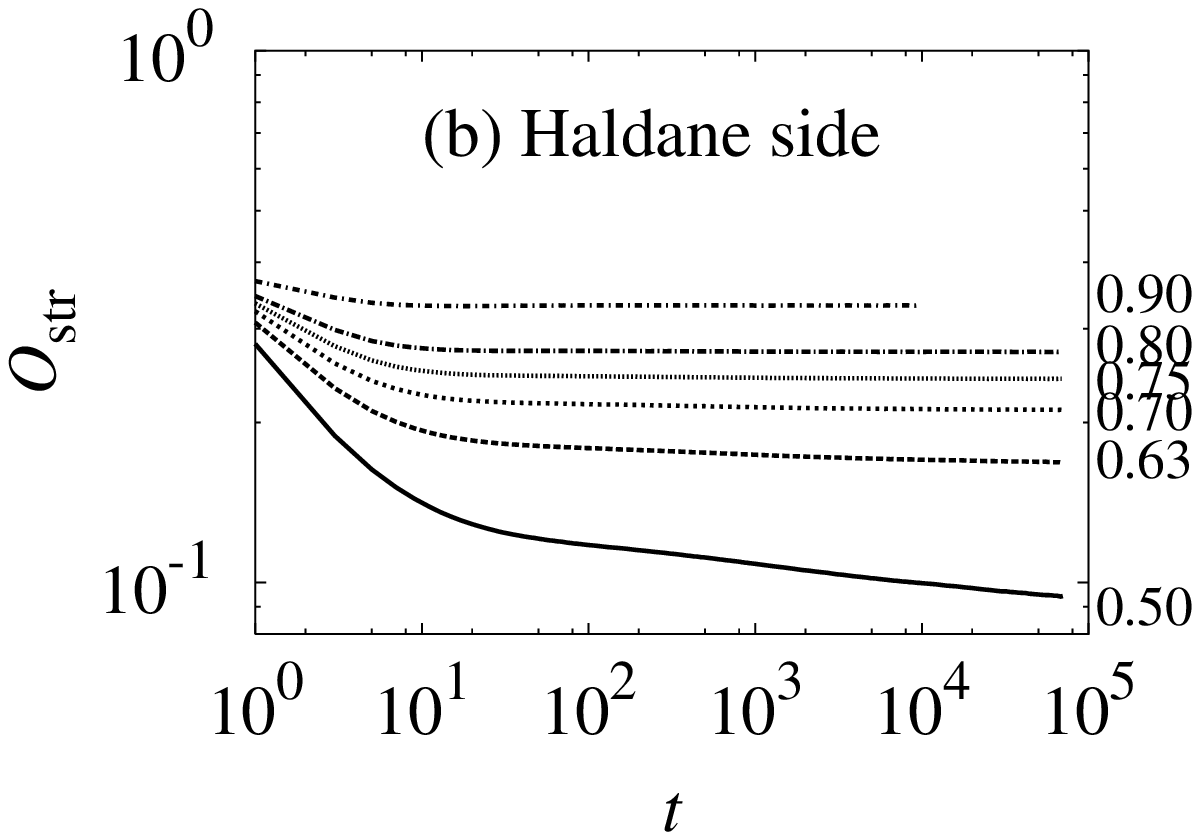}
\end{center}
\caption{Nonequilibrium relaxation of the string order parameters.
A ferromagnetic bond concentration $p$ is denoted beside the data.
}
\label{fig:string}
\end{figure}

Figure \ref{fig:string} shows nonequilibrium relaxation functions of the
string order parameters.
Simulation conditions are the same as those for the local susceptibility.
On the Haldane side it exhibits a converging behavior for $p>0.63$.
This is consistent with our estimate for the gapless transition point
$p=0.625$.
Therefore, the Haldane phase is considered to exist in a region of
$0.625 \le p \le 1$.
This phase is quite robust against randomness.
When the concentration is lowered, the string order parameters decay
algebraically, whose exponent is dependent on the concentration.
On the dimer side the string order parameters suddenly start decaying by an 
introduction of random bonds.
This is also consistent with the behavior of local susceptibility.
The dimer gap state is quite fragile against randomness.

\subsection{The magnetic susceptibility}

\begin{figure}
\begin{center}
\includegraphics[width=8.0cm]{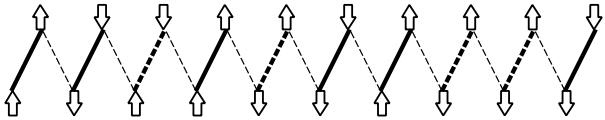}
\end{center}
\caption{A generalized staggered state.
Arrows depict the $S^z$ representation.
Solid lines depict ferromagnetic bonds
and broken lines depict antiferromagnetic bonds. 
}
\label{fig:gstpattern}
\end{figure}

The ground states in pure concentration limits ($p=0, 1$) are 
a spin-disordered state with a finite excitation energy gap.
The magnetic susceptibility takes a value of order unity.
Starting simulation from this ground state, we observe nonequilibrium
relaxation functions of the following two kinds of the magnetic susceptibility.
One is the uniform staggered susceptibility.
This is intended to check the possibility of criticality of the 
uniform antiferromagnetic order speculated on by the experiment of
IPACu(Cl$_x$Br$_{1-x}$)$_3$.
The other one is the generalized staggered susceptibility, 
which is the susceptibility associated with the general staggered state.
Here, we mean the generalized staggered state in Fig. \ref{fig:gstpattern}.
This is a random classical state which minimizes the $S^z$ part of the 
Hamiltonian.
The susceptibility shows converging behavior if the magnetic order remains 
disordered.  It shows diverging behavior if the magnetic order is critical.
In the gapful region estimated in the previous subsections both sets of 
the susceptibility data should exhibit converging behaviors.

\begin{figure}
\begin{center}
\includegraphics[width=8.0cm]{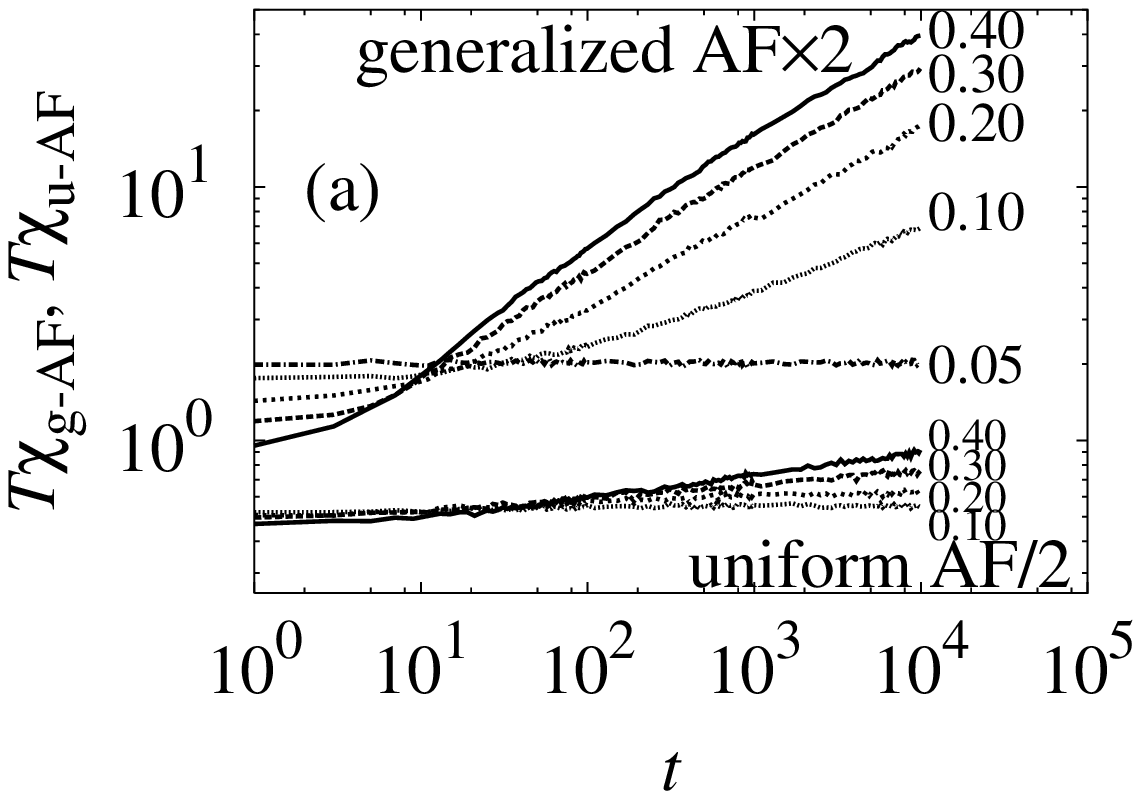}
\includegraphics[width=8.0cm]{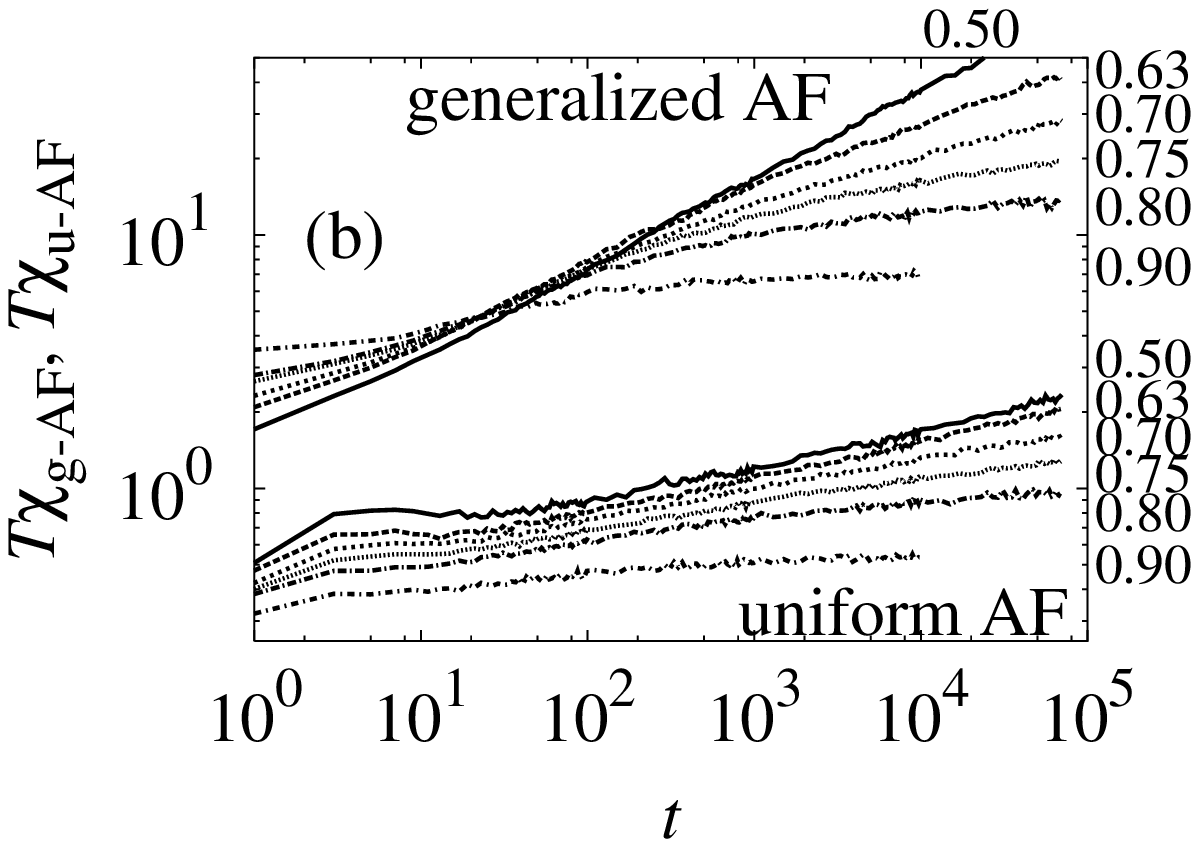}
\end{center}
\caption{
Nonequilibrium relaxation of the generalized staggered susceptibility
$\chi_\mathrm{g-AF}$ and that of the uniform staggered susceptibility
$\chi_\mathrm{u-AF}$. (a) The dimer side. (b) The Haldane side.
Both orders exhibit critical behavior in the gapless region $0 < p < 0.625$.
}
\label{fig:ugst}
\end{figure}

Figure \ref{fig:ugst} shows nonequilibrium relaxation functions of two kinds of 
the magnetic susceptibility.
Simulation conditions are the same as in the previous subsections.
Both sets of the susceptibility data exhibit critical behaviors 
in the gapless region $ 0.2 < p < 0.63$.
There is a grey zone between $p=0.63$ and $p=0.75$ as in the data of the
local susceptibility.
The raw relaxation data alone cannot show a difference between diverging and
converging.
At $p=0.1$ and $p=0.2$ the generalized staggered susceptibility shows
diverging behavior, while the uniform staggered susceptibility remains finite.
We discuss this issue in the next subsection with regard to resolution of
the present simulations.
In the Haldane phase, relaxation functions also exhibit converging behavior.
There is no qualitative difference between the two orders.
Amplitude of the uniform staggered susceptibility is smaller than the
generalized one.
This is because the randomly-located ferromagnetic bonds conflict with the
uniform antiferromagnetic order and this decreases the amplitude.

The critical behavior of the uniform antiferromagnetic order in this figure
is a new finding in the present random model.
It suggests that the order can survive against the randomly located 
ferromagnetic bonds.
This is a purely quantum effect.
We cannot expect it in the classical model.
In the real compound the uniform antiferromagnetic order can be a long-range
order by finite interchain antiferromagnetic couplings.
In order to check the criticality of this order we observe nonequilibrium
relaxation functions of uniform staggered magnetization when we
start simulations from the uniform antiferromagnetic state.

\subsection{Uniform staggered magnetization}

Uniform staggered magnetization ($M_\mathrm{AF}$)
is observed to see whether it decays exponentially or algebraically
when starting from the uniform antiferromagnetic state.
If it exhibits an algebraic decay, this state is proved to be
critical as suggested in the previous subsection.
The system size of the present simulation is $N=1601$ (3202 spins), 
$m=400$, and $\beta J=200$.
The number of random bond configurations is more than one hundred.
The temperature corresponds to $\sim 0.1$ K which is lower than the 
experimental ones (the phase transition was observed at 15 K).
It is also confirmed that there is no temperature dependence 
within the time scale the simulations are performed.
Therefore, we consider that the ground state is realized.
The size of the simulation determines the resolution limit of an
obtained physical quantity.
The resolution of the staggered magnetization is of the order of 
$1/\sqrt{2mN\times {\rm (number~of~samples)}} \le 10^{-4}$.
On the other hand, the resolution of the susceptibility in the previous
subsection is of the order of $J/T$, which
corresponds to the AF magnetization per spin 
$\langle M_{\rm AF} \rangle \sim 1/\sqrt{mN}\sim 3\times 10^{-3}$.
Both simulations are consistent when within the resolution of 
$\langle M_{\rm AF} \rangle > 10^{-3}$.

\begin{figure}
\begin{center}
\includegraphics[width=8.0cm]{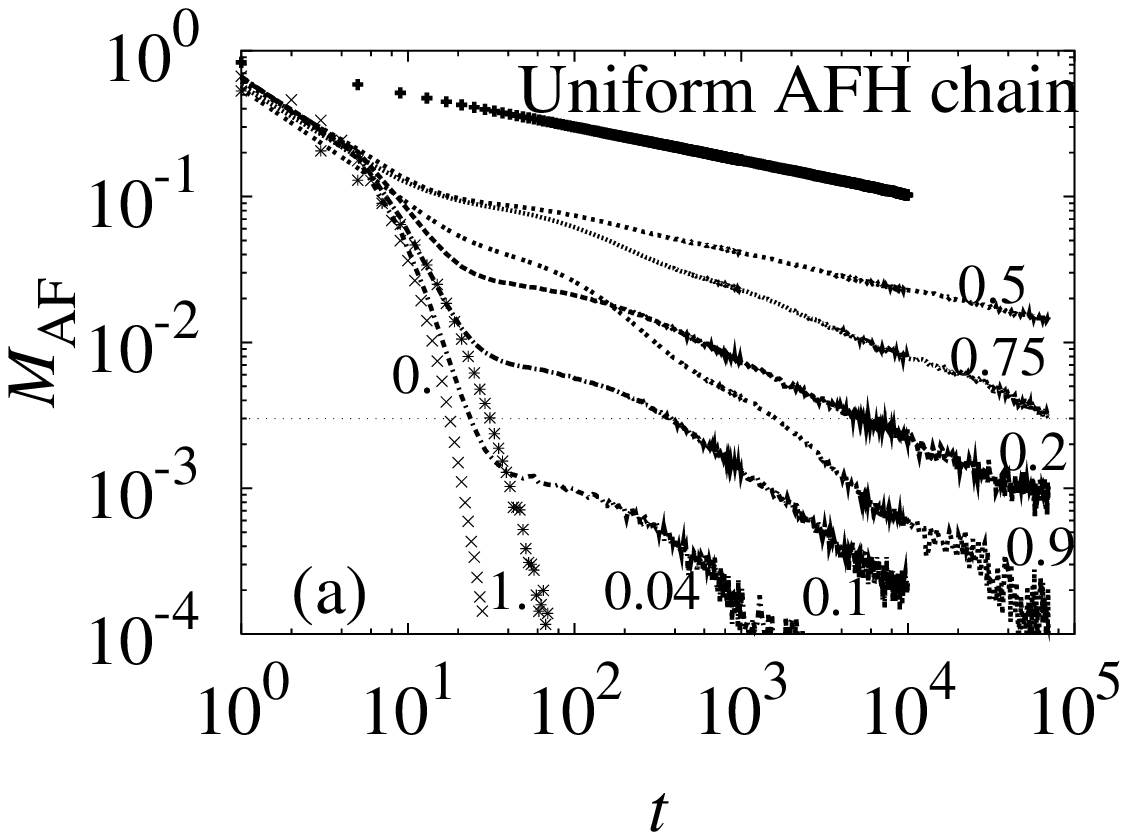}
\includegraphics[width=8.0cm]{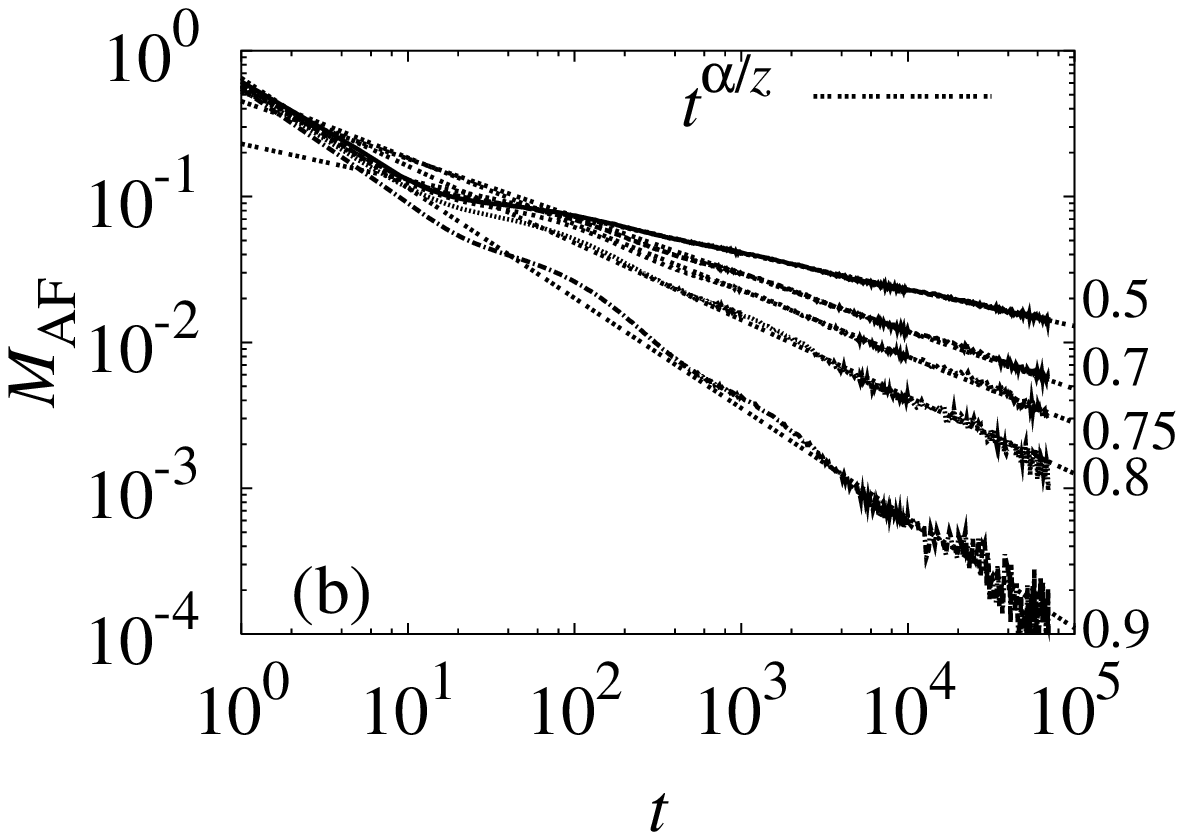}
\end{center}
\caption{(a) Nonequilibrium relaxation of uniform staggered magnetization 
when simulation starts from the uniform AF state.
The value of the ferromagnetic bond concentration $p$
is denoted beside each plot.
Relaxation functions of the pure concentration limit, $p=0$ and $p=1$, and
that of the $S=1/2$ uniform antiferromagnetic Heisenberg (AFH) chain 
are also plotted by symbols.
(b) Relaxation functions on the Haldane side can be fitted by 
$t^{\alpha/z}$, which is analytically obtained by a real-space renormalization
procedure.
}
\label{fig:figaf-fit}
\end{figure}

Figure \ref{fig:figaf-fit} shows nonequilibrium relaxation of staggered
magnetization.
We have also plotted the result of the $S=1/2$ uniform antiferromagnetic
Heisenberg chain, which is exactly solved and the AF order is critical.
In the intermediate region, the relaxation function exhibits algebraic
decay after an initial relaxation that rides on
relaxation of the pure gapful system.
Note that the length of the initial relaxation, $t<50$, 
agrees with that of the susceptibility simulations in Fig. \ref{fig:ugst}.
The relaxation functions of the susceptibility begin diverging behaviors
after 50 Monte Carlo steps.

The numerical value of the staggered magnetization at which the relaxation
begins to exhibit critical behavior is roughly considered
as the magnitude of the order.
At $p=0.1$, an algebraic decay begins below 
the susceptibility resolution limit $3 \times 10^{-3}$.
Therefore, algebraic divergence cannot be observed 
at $p=0.1$ in Fig. \ref{fig:ugst}.
It is difficult to estimate the resolution of the experiment observing the 
magnetic phase transition.
However, if one assumes it to be 
$\langle M_{\rm AF} \rangle \sim 10^{-3}$, the numerical results presented
in this paper quantitatively explain the experimental results.
The experiment can detect the phase transition at $p=0.2$ because the 
amplitude of the magnetic order is $\langle M_{\rm AF} \rangle \sim 10^{-2}$,
where the algebraic decay begins in Fig. \ref{fig:figaf-fit}(a).
This bond concentration corresponds to the Cl atom concentration 
$x=0.45$ by $p=x^2$.
This value coincides with the experimental limit of observing 
the phase transition: $x=0.44$.
At $p=0.1$ the amplitude of the magnetic order is about $10^{-3}$.
It is difficult for the experiment to detect this small magnetic order.
Within the accuracy of Fig. \ref{fig:figaf-fit}(a),
the phase boundary on the dimer side resides between $p=0.04$ and $p=0.1$.
This corresponds to between $x=0.2$ and $x=0.3$, which is lower than the 
experimental observation.
When the resolution of the experimental probe becomes sharper,
the magnetic phase may be observed in a wider region.
The same argument is possible in our simulations.
If resolution of our simulation becomes sharper, the critical behavior can be
observed in a wider region.
A true phase boundary may be located at a concentration lower than $p=0.04$.
The present estimate is the upper limit due to the limited numerical resolution.

For $p > 0.75$, the relaxation shows multi-exponential decay,
suggesting a discrete distribution of the energy gap.
As $p$ decreases, the relaxation appears to be an algebraic decay.
At the fully random point $p=0.5$, the magnetization takes a maximum value.
The slope at this point is almost the same as that of 
the $S=1/2$ uniform AFH spin chain with the amplitude reduced to 1/4.
This is a notable finding in this paper.
The fully random chain exhibits the antiferromagnetic criticality which is
qualitatively equivalent to the nonrandom uniform AFH chain.
This behavior guarantees again the observation of a weak antiferromagnetic
phase transition assisted by weak interchain interactions in the real compound.

\begin{figure}
\begin{center}
\includegraphics[width=8.0cm]{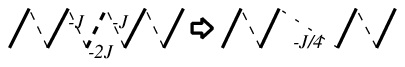}
\end{center}
\caption{Depletion procedure of a strong antiferromagnetic bond on the 
Haldane side.
}
\label{fig:deplete}
\end{figure}

Algebraic decay in the intermediate region
can be explained by the power-law distribution of the energy gap.
On the Haldane side ($p>0.5$), we can consider the strong antiferromagnetic 
bonds as impurities that can be replaced by effective weak bonds.
A singlet dimer state can be easily formed on an isolated strong 
antiferromagnetic bond.
Then, spin degrees of freedom on this bond freeze.
We may deplete this singlet spin pair and connect the neighboring spins by
a new effective interaction bond.
This procedure is depicted in Fig. \ref{fig:deplete}.
A new effective bond is estimated to be $((-J)^2/2*-2J)=-J/4$ 
by the second-order perturbation theory.
As long as successive perturbation is good,
the strength of the effective bond replacing $n$ aligned strong 
antiferromagnetic bonds is $\exp[-\lambda n]J$ with $\lambda=\log 4$.
The probability of the occurrence of this configuration is $(1-p)^n$.
Then, the distribution of the effective bonds is \cite{hida01a,hida01b}
\begin{equation}
 P(J)\sim \frac{1}{\lambda}J^{-1+\frac{1}{\lambda}\log\frac{1}{1-p}}
\propto
\alpha J^{-1+\alpha},
\label{eq:pofj}
\end  {equation}
with $\alpha=(1/\lambda)\log (1/(1-p))$.
Because the Valence-Bond-Solid picture is valid in this region, the singlet
dimers are located on the weak antiferromagnetic bonds between the strong
ferromagnetic bonds.
This bond distribution is directly interpreted by the gap distribution
$P(\Delta)$.
This distribution is equivalent to that in the random singlet phase,
$P(\Delta)=\alpha\Omega^{-\alpha}\Delta^{\alpha-1}$, with a characteristic
energy scale $\Omega$.\cite{fisher94}
Each energy gap contributes to the exponential decay of the
staggered magnetization with a contribution $\exp[-\Delta^z t]$
($z$: the dynamic exponent).
The sum of these exponential decays by the distribution $P(\Delta)$
becomes an algebraic decay as
\begin{equation}
\int \exp[-\Delta^z t]P(\Delta)d\Delta=\Gamma(\alpha/z+1)(t/\tau)^{-\alpha/z},
\label{eq:gamma}
\end  {equation}
with a characteristic time scale $\tau \sim \Omega^{-z}$.
Therefore, the staggered magnetization is expected to decay algebraically with 
the exponent $\alpha/z=(1/z\lambda)\log(1/(1-p))$,
which is dependent on the concentration $p$.
The slope of the algebraic decay in Fig. \ref{fig:figaf-fit} (b)
quantitatively agrees with this expression supposing 
$z=2.2$ for $p>0.6$ and $z=2.0$ for $0.5 \le p< 0.6$.
The value $z=2.2$ is consistent with the two-dimensional classical Ising model.
The value $z=2.0$ means that the dynamics of the Monte Carlo simulation are
governed by pure diffusion.
It suggests a random singlet phase.

On the dimer side ($p<0.5$), interpretation is not straightforward.
It is not good to deplete the strong antiferromagnetic bonds which are the 
majority.
Depletion of the strong ferromagnetic bonds (the Haldane cluster) is possible,
and weak effective bonds of the amplitude $\exp[-\lambda' n]$ \cite{hagiwara90}
may replace them.
We can obtain the same critical behavior by this procedure.
However, the estimated values of the exponent $\alpha/z$ do not 
coincide with the numerical results.

\subsection{Phase diagram}
\label{sec:phasediagram}

\begin{figure}
\begin{center}
\includegraphics[width=8.0cm]{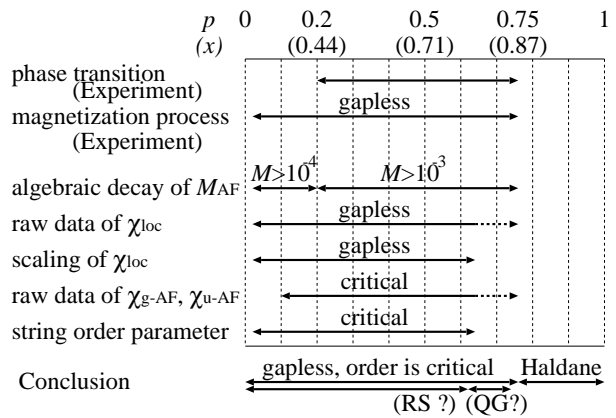}
\end{center}
\caption{
A phase diagram concluded in this paper.
}
\label{fig:phase}
\end{figure}

Putting all the results together we draw a phase diagram 
of the present model in Fig. \ref{fig:phase}.
The corresponding experimental results are also drawn.
The excitation energy gap and the string order parameter remain finite 
in the region $0.75 \le p \le 1$.
This phase boundary $p=0.75$ is a special point.
It is pointed out by Hida \cite{hida99} that the Haldane phase is 
stable as long as the gap distribution is not singular at $\Delta=0$.
It corresponds to $p=0.75$ ($x=0.87$) by $\alpha=1$ in the present model.
The relaxation functions of Fig. \ref{fig:figaf-fit} exhibit multi-exponential
decay for $p > 0.75$.
Therefore, this region is considered the Haldane phase.
The phase boundary of the experimental results is also located at $p=0.75$.
It is very likely that the experiments \cite{manaka3,manakajiba} detected this 
quantum phase transition point.

There is a grey zone between $p=0.625$ and $p=0.75$.
A string order parameter seems to remain finite.
Raw relaxation functions of the local susceptibility suggest that the
excitation gap is zero.
However, scaling analysis suggests that the gap is finite.
Except for the scaling result these lines of evidence suggest this region is
the quantum Griffiths (QG) phase: no excitation gap and a finite string order.
However, classical magnetic orders exhibit critical behavior in this region.
It is not settled whether this phase is the QG phase or not.

It was also reported by Yang and Hyman \cite{yang-h2000}
that the random singlet phase begins at $\alpha\sim 0.67$
for the algebraic bond distribution $P(J)\sim \alpha J^{-1+\alpha}$.
This point corresponds to $p\sim 0.6$.
In the neighborhood of the fully random point, $0.4 < p < 0.6$,
an exponent of the local susceptibility is weakly dependent on $p$, 
suggesting a universal phase.
At the fully random point $p=0.5$, however, relaxation of the uniform 
staggered magnetization is qualitatively the same as the uniform  $S=1/2$
antiferromagnetic Heisenberg chain.
The classical magnetic order is critical in this phase.
It is not clear whether this evidence is compatible with the random singlet
(RS) phase or not.
Therefore, we put a question mark for the RS phase and QG phase in our
phase diagram Fig. \ref{fig:phase}.

\section{Conclusions}
\label{sec:conclusions}
In this paper
the magnetic ordered phase observed in 
the experiment of IPACu(Cl$_x$~Br$_{1-x}$)$_3$ is explained.
The magnetic structure is made clear to be the uniform antiferromagnetic order.
The phase boundaries estimated in the simulation quantitatively agree with
the experimental results.

Behaviors of the relaxation functions in Fig.~\ref{fig:figaf-fit}(a) suggest
the following scenario for the appearance of magnetic order.
Short-range antiferromagnetic correlations are first destroyed by local singlet
dimer states or local Haldane states that appear during the initial relaxation
of exponential decay ($t < 100$).
Beyond these local states, there exist very weak but finite
effective interactions.
This is because local singlet states or local Haldane states are not
the exact eigenstate of the total Hamiltonian.
The effective interactions obey the power-law distribution as in 
Eq. (\ref{eq:pofj}).
They are considered to produce local excitation gaps.
Then, the magnetic order decays exponentially by this excitation gap.
The sum of the exponential decays due to these effective interactions
becomes the algebraic decay as seen in Eq. (\ref{eq:gamma}).
This is considered to be the origin of the criticality of the uniform
antiferromagnetic order. 
Magnetic order is observed in the experiment with the help of 
the interchain interactions.
The present phenomenon is quite interesting because the uniform
antiferromagnetic order survives despite the randomly located 
ferromagnetic bonds.
This may be a new exotic quantum phenomenon.
The role of the ferromagnetic bonds in the quantum system should be 
reconsidered.

It is also noted that the present mechanism explains the
general impurity-induced long-range order phenomenon.
\cite{hase,azuma}
Randomly doped impurities divide a spin chain into local gapful clusters
linked together with small but finite interactions.
The edge spins of the cluster correlate with each other by effective bonds
which exhibit the power-law distribution.
Collections of the correlation become the algebraic decay as in 
Eq. (\ref{eq:gamma}).
Therefore, any classical ordered state can be critical.
In the real compounds, an order whose amplitude is the largest and/or
an order which does not conflict with the interchain interactions
will be selected.

\acknowledgments

The author would like to thank Dr. H.~Manaka and Professor 
K.~Hida for their fruitful discussions and comments.
Use of the random number generator RNDTIK programmed by
Professor N.~Ito and Professor Y.~Kanada is gratefully acknowledged.

\end{document}